\documentclass[%
reprint,
showkeys,
amsmath,amssymb,
aps, physrev,
prl,
]{revtex4-2}

\usepackage[colorlinks = true,
            linkcolor = black!50!red,
            urlcolor  = black!50!red,
            citecolor =  black!50!red,
            anchorcolor = black!50!red]{hyperref}
\usepackage{graphicx,float}
\usepackage{graphicx,float}
\usepackage{xcolor}

\usepackage{bm}
\usepackage{dsfont}
\usepackage{stmaryrd}
\usepackage{tikz}
\usepackage{mdframed}
\usepackage[most]{tcolorbox}
\tcbuselibrary{breakable}
\usepackage{lipsum}
\usepackage[ruled,vlined,linesnumbered]{algorithm2e}
\usepackage{algpseudocode}
\usepackage{pgfplots}

\newcommand{\ds}{\mathrm{d}}
\newcommand{\p}{\partial}
\newcommand{\bb}{\mathbb}
\newcommand{\ff}{\mathfrak}
\newcommand{\cc}{\mathcal}
\RequirePackage[normalem]{ulem} 
\RequirePackage{color}\definecolor{RED}{rgb}{1,0,0}\definecolor{BLUE}{rgb}{0,0,1} 
\providecommand{\DIFaddbegin}{} 
\providecommand{\DIFaddend}{} 
\providecommand{\DIFdelbegin}{} 
\providecommand{\DIFdelend}{} 
\providecommand{\DIFaddbeginFL}{} 
\providecommand{\DIFaddendFL}{} 
\providecommand{\DIFdelbeginFL}{} 
\providecommand{\DIFdelendFL}{} 
\newcommand{\DIFscaledelfig}{0.5}
\RequirePackage{settobox} 
\RequirePackage{letltxmacro} 
\newsavebox{\DIFdelgraphicsbox} 
\newlength{\DIFdelgraphicswidth} 
\newlength{\DIFdelgraphicsheight} 
\LetLtxMacro{\DIFOincludegraphics}{\includegraphics} 
\newcommand{\DIFaddincludegraphics}[2][]{{\color{blue}\fbox{\DIFOincludegraphics[#1]{#2}}}} 
\newcommand{\DIFdelincludegraphics}[2][]{
\sbox{\DIFdelgraphicsbox}{\DIFOincludegraphics[#1]{#2}}
\settoboxwidth{\DIFdelgraphicswidth}{\DIFdelgraphicsbox} 
\settoboxtotalheight{\DIFdelgraphicsheight}{\DIFdelgraphicsbox} 
\scalebox{\DIFscaledelfig}{
\parbox[b]{\DIFdelgraphicswidth}{\usebox{\DIFdelgraphicsbox}\\[-\baselineskip] \rule{\DIFdelgraphicswidth}{0em}}\llap{\resizebox{\DIFdelgraphicswidth}{\DIFdelgraphicsheight}{
\setlength{\unitlength}{\DIFdelgraphicswidth}
\begin{picture}(1,1)
\thicklines\linethickness{2pt} 
{\color[rgb]{1,0,0}\put(0,0){\framebox(1,1){}}}
{\color[rgb]{1,0,0}\put(0,0){\line( 1,1){1}}}
{\color[rgb]{1,0,0}\put(0,1){\line(1,-1){1}}}
\end{picture}
}\hspace*{3pt}}} 
} 
\LetLtxMacro{\DIFOaddbegin}{\DIFaddbegin} 
\LetLtxMacro{\DIFOaddend}{\DIFaddend} 
\LetLtxMacro{\DIFOdelbegin}{\DIFdelbegin} 
\LetLtxMacro{\DIFOdelend}{\DIFdelend} 
\DeclareRobustCommand{\DIFaddbegin}{\DIFOaddbegin \let\includegraphics\DIFaddincludegraphics} 
\DeclareRobustCommand{\DIFaddend}{\DIFOaddend \let\includegraphics\DIFOincludegraphics} 
\DeclareRobustCommand{\DIFdelbegin}{\DIFOdelbegin \let\includegraphics\DIFdelincludegraphics} 
\DeclareRobustCommand{\DIFdelend}{\DIFOaddend \let\includegraphics\DIFOincludegraphics} 
\LetLtxMacro{\DIFOaddbeginFL}{\DIFaddbeginFL} 
\LetLtxMacro{\DIFOaddendFL}{\DIFaddendFL} 
\LetLtxMacro{\DIFOdelbeginFL}{\DIFdelbeginFL} 
\LetLtxMacro{\DIFOdelendFL}{\DIFdelendFL} 
\DeclareRobustCommand{\DIFaddbeginFL}{\DIFOaddbeginFL \let\includegraphics\DIFaddincludegraphics} 
\DeclareRobustCommand{\DIFaddendFL}{\DIFOaddendFL \let\includegraphics\DIFOincludegraphics} 
\DeclareRobustCommand{\DIFdelbeginFL}{\DIFOdelbeginFL \let\includegraphics\DIFdelincludegraphics} 
\DeclareRobustCommand{\DIFdelendFL}{\DIFOaddendFL \let\includegraphics\DIFOincludegraphics} 
\RequirePackage{listings} 
\RequirePackage{color} 
\lstdefinelanguage{DIFcode}{ 
  moredelim=[il][\color{red}\sout]{\%DIF\ <\ }, 
  moredelim=[il][\color{blue}\uwave]{\%DIF\ >\ } 
} 
\lstdefinestyle{DIFverbatimstyle}{ 
	language=DIFcode, 
	basicstyle=\ttfamily, 
	columns=fullflexible, 
	keepspaces=true 
} 
\lstnewenvironment{DIFverbatim}{\lstset{style=DIFverbatimstyle}}{} 
\lstnewenvironment{DIFverbatim*}{\lstset{style=DIFverbatimstyle,showspaces=true}}{} 

\begin{document}

\preprint{APS/123-QED}

\title{From Galactic Clusters to Plasmas in a Single Monte Carlo: \\
Branching Paths Statistics for Poisson-Vlasov/Boltzmann
}

\author{Daniel Yaacoub, St\'ephane Blanco, Richard Fournier,  and Gerjan Hagelaar}

\affiliation{
 UPS, CNRS, INPT, LAPLACE UMR CNRS 5213, Universit\'e de Toulouse,\\
118 route de Narbonne, F-31065 Toulouse, Cedex 9, France
}



\date{\today}

\begin{abstract}
Recent advances have allowed to tackle path-space probabilistic representations of mesoscopic Boltzmann transport nonlinearly coupled to a sub-model of the force-field by step forward approaches in terms of continuous branching stochastic processes.
In this work, path-space probabilistic representations of free-space Poisson-Vlasov and Poisson-Boltzmann systems are exhibited.
This yields novel propagator representations and opens new routes for efficient and reference simulations by use of new branching backward Monte Carlo algorithms.
Subsequent statistical estimator are benchmarked on gravitational clusters and plasmas dynamics.
\end{abstract}


\maketitle
\section{Introduction}

\paragraph{context.}

Poisson-Vlasov or Poisson-Boltzmann dynamics are commonly-used models for mesoscopic transport coupled to a self-consistent force-field model either in gravitational dynamics or in electric transport. Among these systems one can mention:
\begin{itemize}
\item[-] Collisionless or collisional electrostatic plasmas. Collective modes such as Langmuir waves, sheath turbulence, or beam-plasma insabilities are modeled by the self-consistent evolution of the one-particle distribution function coupled to Poisson's equation for the electric potential. Collisions with neutrals can also be added in the main transport equation.
\item[-] Dusty plasmas, planetary rings and charged granular media. Systems composed of charged dust grains or macroscopic particulates embedded in a plasma (or interacting via long-range forces) are frequently modeled with kinetic mean-field descriptions. Vlasov–Poisson variants and reduced kinetic models appear as usefull models \cite{Horanyi_1996}.
\item[-] Electron transport in semiconductors. Semiclassical models for charge transport in semiconductor devices use the (linearized) Boltzmann equation for carrier dynamics coupled self-consistently to Poisson’s equation for the electrostatic potential. These models capture hot-electron effects and nonlocal transport beyond drift–diffusion approximations.
\item[-] Stellar dynamics and large-scale self-gravitating systems. Collisionless stellar systems such as galaxies, globular clusters and large-scale dark-matter structures before shell-crossing are ofently described by Vlasov's equation coupled to Poisson’s one for the gravitational potential. This Poisson-Vlasov description underlies many problems in galactic dynamics, stability theory and the statistical mechanics of self-gravitating systems \cite{Binney_2008,Chavanis_2002}.
\item[-] Cosmology and dark-matter large-scale structure. Cold dark matter on cosmological scales behaves, to first approximation, as a collisionless self-gravitating medium. Its evolution in an expanding background is often modeled by a cosmological Vlasov–Poisson system \cite{Rampf_2021}. This framework is used in analytic studies and in kinetic numerical simulations of structure formation.
\end{itemize}

In the area of plasma physics, one of the most ambitious issue concerns today the effort to design carbon-free power production using magnetic confinement fusion.
Among the challenges on the path to fusion reactors, the management of heat exhaust is one of the most critical ones.
The approach to manage the extreme heat fluxes to the reactor wall relies on the dissipation of the plasma’s energy through interaction with the neutral gas present in the edge of the plasma due to plasma-surface interaction.
The physics at play consists in a balance between plasma transport, dominated by turbulence, and molecular reactions related to plasma-neutrals collisions.
Hence, understanding and predicting such strongly nonlinear dynamics is of a prime interest for the design and operational space definition of future devices.\\


In the area of cosmology and astrophysics, gravitational dynamics are of prime interest since they give access to fundamental aspects of the universe.
In deed, large-scale structures for instance composed by globular clusters or galaxies encode within their own dynamics, set by the gravitation, signatures inherited from an early inflationary period.
To extract this information from observations of the large-scale mass density and velocity distributions of galaxies or other astronomical tracers, one have to predict their temporal dynamics with respect to cosmic time.
Due to the high-dimensional complexity and the nonlinear nature of the underlaying physics, such predections result in a very challenging problem.\\


\paragraph{mathematical model.}

Vlasov's equation consists in the 6D spatio-temporal collisionless evolution equation of particles submitted to a conservative force field $\mathbf{F}=-\boldsymbol{\nabla}\phi$ ; $\phi$ being the potential energy.
Theses particules are described through their one-particle distribution function $f$ at a given phase-space location $(\mathbf{r},\mathbf{c})$ and a given time $t$.
When absorption/killing, creation or scattering events occur, Vlasov's equation turns into the linear Boltzmann equation with an appropriate collision term. The system \eqref{eq:Poisson-Vlasov} depicts the coupling between the free-space Boltzmann transport equation of $f$ and Poisson's equation, standing as a self-consistent submodel of the potential energy $\phi$, for all $\mathbf{r}\in\bb{R}^3$, $\mathbf{c}\in\bb{R}^3$, $t>t_\text{o}$ and given the initial condition $f(\mathbf{r},\mathbf{c},t_\text{o})=f_\text{o}(\mathbf{r},\mathbf{c})$.
\begin{widetext}
\begin{equation}
\left\{
\begin{array}{ll}
\p_tf(\mathbf{r},\mathbf{c},t)&\hspace{-0.15cm}+\mathbf{c}\cdot\boldsymbol{\nabla}_\mathbf{r}f(\mathbf{r},\mathbf{c},t)-(1/m)\boldsymbol{\nabla}_\mathbf{r}\phi(\mathbf{r},t)\cdot\boldsymbol{\nabla}_\mathbf{c}f(\mathbf{r},\mathbf{c},t)=-\nu_\text{e}f(\mathbf{r},\mathbf{c},t)+\nu_\text{a}f^\star(\mathbf{r},\mathbf{c},t)+\nu_\text{d}\int_{\bb{S}^2}\ds\boldsymbol{\omega'}~\varphi(\boldsymbol{\omega}|\boldsymbol{\omega'})f(\mathbf{r},||\mathbf{c}||\boldsymbol{\omega'},t)\\
\nabla^2\phi(\mathbf{r},t)&=\kappa\left(\int_{\bb{R}^3}\ds\mathbf{c}~f(\mathbf{r},\mathbf{c},t)-\rho_\text{ext}(\mathbf{r},t)\right)\\
f(\mathbf{r},\mathbf{c},t_\text{o})&=f_\text{o}(\mathbf{r},\mathbf{c})
\end{array}
\right.\label{eq:Poisson-Vlasov}
\end{equation}
\end{widetext}
The first right-hand side collision term of the transport equation stands for a loss term with the corresponding extinction frequency $\nu_\text{e}$.
The second one stands for a prescribed source term $f^\star$ with the corresponding  absorption frequency $\nu_\text{a}$.
Finally, the last term stands for linear scattering effects given the diffusion frequency $\nu_\text{d}$ and the scattering phase-function $\varphi(\boldsymbol{\omega}|\boldsymbol{\omega'})$.
Concerning Poisson's equation, in the case of self-gravitating structures, $f$ stands for the phase-space distribution function of glaxies, stars or other tracers, $\phi\equiv m\Phi_\text{g}$ and $\kappa=4\pi mG$ ; $G$ being the gravitaional constant.
In the case of charges transported within plasmas, $f$ stands for the phase-space distribution function of electrons or ions, $\phi\equiv q\Phi_\text{e}$ and $\kappa=-q^2/\varepsilon_\text{o}$ ; $\varepsilon_\text{o}$ being the vacuum dielectric premittivity and $q$ the electric charge. \\


\paragraph{numerical simulation methods.}

In plasma physics, cosmology, and astrophysics, the numerical solution of the Poisson–Vlasov system is a central issue for the study of strongly nonlinear dynamics. 
To this aim, Poisson-Vlasov equations are currently solved numerically by N-body approaches such as Particle-In-Cell (PIC) and Particle-Mesh (PM) algorithms \cite{Buneman_1959,Dawson_1983,Morse_1969}. 
In these methods, the phase-space distribution function is represented by an ensemble of particles, that is, a set of Dirac distributions in phase space interacting with each other through self-consistent electric or gravitational forces. 
The particle charge or mass is deposited onto a spatial grid, and Poisson's equation is usually solved on that mesh before the resulting fields are interpolated back to the particle positions to advance their trajectories. 
By replacing direct particle-particle interactions with a field solve on a mesh, these methods significantly reduce the computational cost, remaining yet huge. 
In the same veine, classical N-body methods compute gravitational or electrostatic forces through direct summation or through hierarchical accelerations such as tree or fast multipole algorithms, and are widely used in stellar dynamics and cosmological simulations \cite{Dolag_2008,Colombi_2001}. 
Another important class of solvers consists of grid-based Vlasov methods, among which the semi-Lagrangian splitting scheme of Cheng and Knorr is particularly emblematic \cite{Cheng_1976}. 
In this approach, the phase-space distribution function is discretized on a mesh and advanced by following characteristics backward in time, typically through successive transport steps in configuration and velocity space combined with interpolation. 
Such methods exploit the underlying Liouville structure of the Vlasov equation and remain especially attractive in the warm regime, where the initial velocity dispersion is non-negligible. However, in all cases, the six-dimensional phase-space description of the dynamics leads to substantial computational and algorithmic costs.\\

\paragraph{need for physical clarity and computational feasability.}

When a self-consistent submodel of the force field is involved, that is, when the force itself depends on the transported distribution function, physically insightful probabilistic representations are still largely lacking. 
This coupling is intrinsically nonlinear, even when both the transport equation and the force-field submodel are linear when considered separately. 
Recent advances \cite{Yaacoub_2025} have allowed to tackle path-space probabilistic representation of macroscopic drift-diffusion models nonlinearly coupled to a sub-moedl of the drift velocity field.
In this letter, we aim at recasting such a breaktrhough in mesoscopic transport physics, based on the fact that force-fields or accelerations play the role of drift or advection fields in the velocity-space rather than the usual geometric-space of macroscopic transport models. 
Beyond the conceptual gain, such an approach would be meshless, naturally highly parallelizable, and would inherit the full power of Monte Carlo methods: statistical estimators with confidence intervals, no spatial discretization, and a direct separation between the probabilistic computation and the underlying geometry. 
In this perspective, building scalable methods that simultaneously provide physical clarity and computational feasibility remains a central challenge for both theoretical and applied communities.\\

\section{Branching Path-space probabilistic representation}



\begin{widetext}
\begin{tcolorbox}[
  enhanced,
  breakable,
  width=\textwidth,
  boxrule=0.4pt,
  left=3pt,
  right=3pt,
  colback=gray!10,
  colframe=black
]
  \begin{center}
  \textbf{Force-field probabilistic representation}
  \end{center}
  
$~~~~$Described by Poisson's equation, taken appart form the Vlasov/Boltzmann transport equation, the potential energy $\phi$ undergoes a fully linear and diffusive model.
Hence, Feynman-Kac probabilistic representation of $\phi$ \cite{Kakutani_1944,Feynman_1948,Kac_1951,Wiener_1921}
\begin{equation}
\phi(\mathbf{r},t)=\bb{E}_{\boldsymbol{\cc{R}}_s^\phi}\left[-\kappa\int_\text{o}^{+\infty}\hspace{-0.3cm}\ds s\left(\int_{\bb{R}^3}\ds\mathbf{c}~f\big(\boldsymbol{\cc{R}}_s^\phi,\mathbf{c},t\big)-\rho_\text{ext}\big(\boldsymbol{\cc{R}}_s^\phi,t\big)\right)\right]
\end{equation}
holds and is supported by the standard brownian process $\{\boldsymbol{\cc{R}}^\phi_s\}_s$ defined by the stochastic differential equation $\ds\boldsymbol{\cc{R}}_s^\phi=\sqrt{2}\ds\boldsymbol{\ff{W}}_s$ and the initial condition $\boldsymbol{\cc{R}}_\text{o}^\phi=\mathbf{r}$.
Starting from this path-space probabilistic representation of $\phi$,one can easily deduce the Feynman-Kac's represention of $\boldsymbol{\nabla}_\mathbf{r}\phi$.
In deed, since the initial value nonlinear transport probelm depicts a free-space problem, Malliavin stochastic calculus provides us with a direct differentiation rule:
\begin{equation}
\boldsymbol{\nabla}_\mathbf{r}\phi(\mathbf{r},t)=\bb{E}_{\boldsymbol{\cc{R}}_s^\phi}\Bigg[-\kappa\int_\text{o}^{+\infty}\hspace{-0.3cm}\ds s\left(\int_{\bb{R}^3}\ds\mathbf{c}~f\big(\boldsymbol{\cc{R}}_s^\phi,\mathbf{c},t\big)-\rho_\text{ext}\big(\boldsymbol{\cc{R}}_s^\phi,t\big)\right)\left(\frac{\boldsymbol{\cc{R}}_s^\phi-\mathbf{r}}{2s}\right)\Bigg]\label{eq:grad1}
\end{equation}
involving the first variation stochastic process, identical to $\{\boldsymbol{\cc{R}}_s^\phi\}_s$ in the case of standard brownian processes, and Malliavin's weight $(\boldsymbol{\cc{R}}_s^\phi-\mathbf{r})/2s$ \cite{Bismut_1984,Elworthy_1994,Fournie_1999,Chen_2007} which can be seen as an equivalent form of $\boldsymbol{\nabla}_\mathbf{r}\text{ln}(p_{\boldsymbol{\cc{R}}_s^\phi}(\boldsymbol{\cc{R}}_s^\phi,s|\mathbf{r},\text{o}))$ \cite{Lecuyer_1990, Lecuyer_1991,Warren_2012}.
Finally, nested integrals over time and velocity-space can themself be interpreted as expectations over random variables.
By introducing importance probability distribution functions $\ds\bb{P}\{\cc{S}^\phi=s\}=p_\cc{S^\phi}(s)\ds s$ and $\ds\bb{P}\{\mathbf{C}=\mathbf{c}\}=p_\mathbf{C}(\mathbf{c})\ds \mathbf{c}$, equation \eqref{eq:grad1} can be read as
\begin{equation}
\boldsymbol{\nabla}_\mathbf{r}\phi(\mathbf{r},t)=\bb{E}_{\cc{S}^\phi,\mathbf{C},\boldsymbol{\cc{R}}_\cc{S^\phi}^\phi}\Bigg[\left(
\rho_\text{ext}\big(\boldsymbol{\cc{R}}_{\cc{S}^\phi}^\phi,t\big)-\frac{f\big(\boldsymbol{\cc{R}}_{\cc{S}^\phi}^\phi,\mathbf{C},t\big)}{p_{\mathbf{C}}(\mathbf{C})}\right)\frac{\kappa(\boldsymbol{\cc{R}}_\cc{S^\phi}^\phi-\mathbf{r})}{2\cc{S^\phi}p_\cc{S^\phi}(\cc{S}^\phi)}\Bigg]\label{eq:grad2}
\end{equation}
Admitting for now that $f(\mathbf{r},\mathbf{c},t)$ can be represented as an expectation $\bb{E}_{\cc{F}}[\cc{F}|\mathbf{r},\mathbf{c},t]$, as we will show in the next paragraph, equation \eqref{eq:grad2} ultimately leads to the path-space Feynman-Kac probabilistic representation 
\begin{equation}
\boldsymbol{\nabla}_\mathbf{r}\phi(\mathbf{r},t)=\bb{E}_{\cc{S}^\phi,\mathbf{C},\boldsymbol{\cc{R}}_\cc{S}^\phi,\cc{F}}\Bigg[\left(
\rho_\text{ext}\big(\boldsymbol{\cc{R}}_{\cc{S}^\phi}^\phi,t\big)
-\frac{\big(\cc{F}\big|\boldsymbol{\cc{R}}_{\cc{S}^\phi}^\phi,\mathbf{C},t\big)}{p_{\mathbf{C}}(\mathbf{C})}\right)\frac{\kappa(\boldsymbol{\cc{R}}_{\cc{S}^\phi}^\phi-\mathbf{r})}{2\cc{S}^\phi p_{\cc{S}^\phi}(\cc{S}^\phi)}\Bigg]\equiv\bb{E}_{\boldsymbol{\cc{G}}_\phi}\left[\boldsymbol{\cc{G}}^\phi\big|\mathbf{r},t\right]\label{eq:grad3}
\end{equation}
of $\boldsymbol{\nabla}_\mathbf{r}\phi$.

\end{tcolorbox}
\end{widetext}


\begin{widetext}
\begin{tcolorbox}[
  enhanced,
  breakable,
  width=\textwidth,
  boxrule=0.4pt,
  left=3pt,
  right=3pt,
  colback=gray!10,
  colframe=black
]
  \begin{center}
  \textbf{Path-space probabilistic representation}
  \end{center}
  
$~~~~$In the case of a prescribed force-field $\mathbf{F}(\mathbf{r},t)=-\boldsymbol{\nabla}_\mathbf{r}\phi(\mathbf{r},t)$, probabilistic path-space representations of the distribution function $f$ can be exhibited starting from the integral formulation of Boltzmann transport equation \eqref{eq:Poisson-Vlasov} \cite{Maire_2006,Lejay_2010}:
\begin{equation}
f(\mathbf{r},\mathbf{c},t)=f_\text{o}\big(\boldsymbol{\cc{R}}_{t-t_\text{o}},\boldsymbol{\cc{C}}_{t-t_\text{o}}\big)\text{e}^{-\nu_\text{e}(t-t_\text{o})}+\int_\text{o}^{t-t_\text{o}}\hspace{-0.35cm}\ds s\left(\nu_\text{a}f^\star(\boldsymbol{\cc{R}}_s,\boldsymbol{\cc{C}}_s,t-s)+\nu_\text{d}\int_{\bb{S}^2}\hspace{-0.15cm}\ds\boldsymbol{\omega'}~\varphi(\boldsymbol{\omega}|\boldsymbol{\omega'})f(\boldsymbol{\cc{R}}_s,||\boldsymbol{\cc{C}_s}||\boldsymbol{\omega'},t-s)\right)\text{e}^{-\nu_\text{e}s}\label{eq:integral}
\end{equation}
given the initial distribution function value $f_\text{o}$.
Ballistic paths $\{(\boldsymbol{\cc{R}}_s,\boldsymbol{\cc{C}}_s)\}_s$ are solutions of $\ds\boldsymbol{\cc{R}}_s=-\boldsymbol{\cc{C}}_s\ds s$ and $\ds\boldsymbol{\cc{C}}_s=(\boldsymbol{\nabla}_\mathbf{r}\phi(\boldsymbol{\cc{R}}_s,s)/m)\ds s$, and bacwardly propagate initial/volumic sources in the sense of Green starting from the probe position $(\boldsymbol{\cc{R}}_\text{o},\boldsymbol{\cc{C}}_\text{o})=(\mathbf{r},\mathbf{c})$.
$f$ thus results in the exponentially attenuated sources encountered along paths.\\

$~~~~$ A subsequent representation of $f$ can be derived in terms of survival time $\cc{S}$ from the integral formulation.
In deed, 
\begin{equation}
f(\mathbf{r},\mathbf{c},t)=\int_\text{o}^{+\infty}\ds s~\nu_\text{e}\text{e}^{-\nu_\text{e}s}\left[\begin{matrix}\mathds{1}_{\{s\geq t\}}f_\text{o}\big(\boldsymbol{\cc{R}}_{t-t_\text{o}},\boldsymbol{\cc{C}}_{t-t_\text{o}}\big)\\
+\mathds{1}_{\{s< t\}}\left(\frac{\nu_\text{a}}{\nu_\text{e}}f^\star(\boldsymbol{\cc{R}}_s,\boldsymbol{\cc{C}}_s,t-s)+\frac{\nu_\text{d}}{\nu_\text{e}}\int_{\bb{S}^2}\ds\boldsymbol{\omega'}~\varphi(\boldsymbol{\omega}|\boldsymbol{\omega'})f(\boldsymbol{\cc{R}}_s,||\boldsymbol{\cc{C}}||\boldsymbol{\omega'},t-s)\right)\end{matrix}\right]\label{eq:duhamel}
\end{equation}
Introducing now the probability density function $\ds\bb{P}\{\cc{S}=s\}=\nu_\text{e}\text{e}^{-\nu_\text{e}s}\ds s$ and interpreting the scattering phase function $\varphi(\boldsymbol{\omega}|\boldsymbol{\omega'})$ the probability density function $\ds\bb{P}\{\boldsymbol{\Omega'}=\boldsymbol{\omega'}\}=\varphi(\boldsymbol{\omega}|\boldsymbol{\omega'})\ds\boldsymbol{\omega'}$, equation \eqref{eq:duhamel} can be written 
\begin{equation}
\begin{split}
f(\mathbf{r},\mathbf{c},t)&=\bb{E}_{\cc{S},\cc{B},\boldsymbol{\Omega'}}\left[\mathds{1}_{\{\cc{S}\geq t\}}f_\text{o}\big(\boldsymbol{\cc{R}}_{t-t_\text{o}},\boldsymbol{\cc{C}}_{t-t_\text{o}}\big)+\mathds{1}_{\{\cc{S}< t\}}\left(\cc{B}f^\star(\boldsymbol{\cc{R}}_\cc{S},\boldsymbol{\cc{C}}_\cc{S},t-\cc{S})+(1-\cc{B})f(\boldsymbol{\cc{R}}_\cc{S},||\boldsymbol{\cc{C}}_\cc{S}||\boldsymbol{\Omega'},t-\cc{S})\right)\right]\\
&\equiv\bb{E}_{\cc{F}}[\cc{F}|\mathbf{r},\mathbf{c},t]
\end{split}
\end{equation}
given, Bernoulli's random variable $\cc{B}$ of probability $\bb{P}_\text{a}=\nu_\text{a}/\nu_\text{e}$.
Finally, by the linearity property of expectations, one concludes
\begin{equation}
f(\mathbf{r},\mathbf{c},t)=\bb{E}_{\cc{S},\cc{B},\boldsymbol{\Omega'},\cc{F}}\left[\mathds{1}_{\{\cc{S}\geq t\}}f_\text{o}\big(\boldsymbol{\cc{R}}_{t-t_\text{o}},\boldsymbol{\cc{C}}_{t-t_\text{o}}\big)+\mathds{1}_{\{\cc{S}< t\}}\left(\cc{B}f^\star(\boldsymbol{\cc{R}}_\cc{S},\boldsymbol{\cc{C}}_\cc{S},t-\cc{S})+(1-\cc{B})(\cc{F}|\boldsymbol{\cc{R}}_\cc{S},||\boldsymbol{\cc{C}}_\cc{S}||\boldsymbol{\Omega'},t-\cc{S})\right)\right]\label{eq:Esp}
\end{equation}

The solution of Poisson-Vlasov/Boltzmann transport equation \eqref{eq:Poisson-Vlasov} can be then conceptualized in terms of an expectation over a stochastic process in the vein of Feynman-Kac probabilistic representations.
Whenever the force-field $\mathbf{F}$ is prescribed, ballistic path are fully deterministic between two events such as scattering or absorption.\\

$~~~~$However, in the case of a nonlinear force-field coupling as it is for the Poisson-Vlasov/Boltzmann system \eqref{eq:Poisson-Vlasov}, $\mathbf{F}$ depends itself on the own solution $f$ of the main transport model through Poisson's self-consistent sub-model.
In the latter case, recente advances make possible to build path-space probabilistic pictures without knowing explicitely and independently the prescibed force field $\mathbf{F}(\mathbf{r},t)$ but having instead a probabilistic representation $\mathbf{F}(\mathbf{r},t)=-\bb{E}_{\boldsymbol{\cc{G}}^\phi}[\boldsymbol{\cc{G}}^\phi|\mathbf{r},t]$ of the sub-model and knowing only the statistics of $\boldsymbol{\cc{G}}^\phi$.
The two perspectives of representations presented in \cite{Yaacoub_2025} are hereafter briefly exposed: \textbf{1.} The usual McKean representation seen as continuously inlaying the full path-space representation $\mathbf{F}=-\bb{E}_{\boldsymbol{\cc{G}}^\phi}[\boldsymbol{\cc{G}}^\phi]$ within ballitic paths involved between two events. \textbf{2.} The coupled path-space representation allowing to drive ballistic paths between two events without knowing $\mathbf{F}$ but knowing instead only the statistics of $\boldsymbol{\cc{G}}^\phi$.\\

  \begin{minipage}[t]{0.49\textwidth}
    \textbf{1. McKean-Feynman-Kac inlaid representation.}\\
    
$~~~~$McKean representation reads as
\begin{equation}
\left\{
\begin{array}{ll}
\ds\boldsymbol{\cc{R}}_s&=-\boldsymbol{\cc{C}}_s\ds s\\
\ds\boldsymbol{\cc{C}}_s&=(1/m)\bb{E}[\boldsymbol{\cc{G}}^\phi|\boldsymbol{\cc{R}}_s,t-s]\ds s
\end{array}
\right.\label{eq:McKean}
\end{equation}
Since $\boldsymbol{\nabla}_\mathbf{r}\phi(\mathbf{r},t)=\bb{E}_{\boldsymbol{\cc{G}}^\phi}[\boldsymbol{\cc{G}}^\phi|\mathbf{r},t]$, it obviously allows to recover deterministic balistic paths between two events - collision, scattering, killing -.
At each time $s\in[\text{o},t-t_\text{o}]$ the knowledge of this McKean
paths $\{(\boldsymbol{\cc{R}}_s,\boldsymbol{\cc{C}}_s)\}_s$ implies the one of $\bb{E}_{\boldsymbol{\cc{G}}^\phi}\left[\boldsymbol{\cc{G}}^\phi|\boldsymbol{\cc{R}}_{s'},t-s'\right]$ for all $s'<s$, \textit{i.e.} the whole force-field map.
A path is constructed by inlaying a full force-field path-space centered at each $(\boldsymbol{\cc{R}}_{s'},\boldsymbol{\cc{C}}_{s'})$, drawing then, an infinite tree of inlaid path-spaces.
The cost is huge since statistical estimations based on this formulation either by particle-systems approaches or by pointwise Monte Carlo methods present a computational time explosion besides the loss of being able to define a unique branching stochastic process.

  \end{minipage}\hfill
  \begin{minipage}[t]{0.49\textwidth}
    \textbf{2. Coupled Feynman-Kac representation.}\\
    
$~~~~$Our coupled representation reads as
\begin{equation}
\left\{
\begin{array}{ll}
\ds\boldsymbol{\widetilde{\cc{R}}}_s&=-\boldsymbol{\widetilde{\cc{C}}}_s\ds s\\
\ds\boldsymbol{\widetilde{\cc{C}}}_s&=(1/m)(\boldsymbol{\cc{G}}^\phi|\boldsymbol{\widetilde{\cc{R}}}_s,t-s)\ds s
\end{array}
\right.\label{eq:BSP}
\end{equation}
Between two events - either collisions, killing, or scattering - ballistic paths become then stochastic because of the random acceleration appearing now in eqn.\eqref{eq:BSP}. Such paths are described by the embedded phase-space stochastic process $\{(\boldsymbol{\widetilde{\cc{R}}}_s,\boldsymbol{\widetilde{\cc{C}}}_s)\}_s$ backwardly propagating toward sources.
At each time $s\in[\text{o},t-t_\text{o}]$, the knowledge of this process is now entirely determined by $\boldsymbol{\cc{G}}^\phi|\widetilde{\boldsymbol{\cc{R}}}_{s'},t-s'$ for all $s'<s$, that is the statistics of $\boldsymbol{\cc{G}}^\phi$ only, in contrast with the full force-field that was required above. 
A path $\{\tilde{\mathbf{r}}_s,\tilde{\mathbf{c}}_s\}_s$ is constructed by embedding a unique path of $\boldsymbol{\cc{G}}^\phi$ centered at each $(\mathbf{r}_{s'},\mathbf{c}_{s'})$.
In other words, force-field paths ($\boldsymbol{\cc{G}}$-paths) pass on all the
information about the coupled force-field model, without having to inlay a full force-field path-space but drawing instead a unique branch.
$\{\widetilde{\boldsymbol{\cc{R}}}_s,\widetilde{\boldsymbol{\cc{C}}}_s\}_s$ can therefore be understood as an embedded process that includes the statistics of $\boldsymbol{\cc{G}}^\phi$, and thus recast our formulation within Feynman-Kac's theoretical framework \cite{Feynman_1948,Kac_1949}.
As introduced in \cite{Yaacoub_2025}, this counterintuitive viewpoint is constructed as the continuous limit of a branching process, allowing an exact probabilistic representation without having to know the whole force-field $-\boldsymbol{\nabla}_\mathbf{r}\phi=-\bb{E}[\boldsymbol{\cc{G}}^\phi]$ everywhere, but being accelerated instead by a random force-field $-\boldsymbol{\cc{G}}^\phi$.

  \end{minipage}
\end{tcolorbox}
\end{widetext}

\section{Monte Carlo method and statistical estimates}

backward, pointwise, meshfree

\paragraph{Monte Carlo method.}

Starting now from the probabilistic representation \eqref{eq:Esp}-\eqref{eq:BSP}, the Monte Carlo method allows us to build the following statistical estimator 
\begin{equation}
\widehat{F}_N(\mathbf{r},\mathbf{c},t)=\frac{1}{N}\sum_{i\in\llbracket1;N\rrbracket}(\cc{F}_i|\mathbf{r},\mathbf{c},t)\label{eq:estimator}
\end{equation}
based on the family of independent random variables $\{\cc{F}_i\}_{i\in\llbracket 1;N\rrbracket}$, of corresponding realizations $\mathfrak{f}_i$, and identically distributed with rescpect to $\cc{F}$.
As the number $N$ of samples tend to infinity, $\widehat{F}_N(\mathbf{r},\mathbf{c},t)$ converges in probability toward $f(\mathbf{r},\mathbf{c},t)$ by the law of large numbers.
Algorithm \ref{alg:MC} depicts how to build a pointwise statistical estimation of distribution function $f$ at a given phase-space probe position $(\mathbf{r},\mathbf{c},t)$ by use of the statistical estimator \eqref{eq:estimator}.

\begin{algorithm}[H]
\label{alg:MC}
  \caption{Single Branching path-space Monte Carlo}
  \label{}
   $\bullet$ number of realisations : $N$\;
   $\bullet$ probe position : $(\mathbf{r},\mathbf{c},t)$\;
   $\bullet$ initialisation : $i=0$, $\Sigma=0$, $\Sigma_2=0$\;
   \While{$i<N$}{
   $\bullet$ $f_i\leftarrow$ sample a distribution function random variable $\cc{F}_i$ starting at $(\mathbf{r},\mathbf{c},t)$ according to Alg. \ref{alg:function}\;
   $\bullet$ $\Sigma\leftarrow\Sigma+\mathfrak{f}_i$\;
   $\bullet$ $\Sigma_2\leftarrow\Sigma_2+\mathfrak{f}_i^2$\;
   $\bullet$ $i\leftarrow i+1$\;}
   \Return statistiacal estimation: $\Sigma/N$\\
   \Return statistical standard deviation: $\sqrt{(\Sigma_2/N-(\Sigma/N)^2)/(N-1)}$
\end{algorithm}
Along with Alg. \ref{alg:function} and Alg. \ref{alg:force}, Alg. \ref{alg:MC} provides a statistical sampling of the unique path-space underlaying the exact probabilistic representation \eqref{eq:Esp}-\eqref{eq:BSP}.
The corresponding paths are branching ones but not paths inlaid with a full path-space, as it would be for McKean representation.
The latter would lead to nesting a Monte Carlo estimations within Monte Carlo estimations, as it is done in the context of macroscopic velocity-coupled models by recent works in the community of computer graphics \cite{Rioux_2022,Sugimoto_2024}.

Te BBMC stastistical estimation procedure presented in this section benefits from all the power of usual pointwise and path-space Monte Carlo algorithms.
As $\delta s\to 0$, the statistical estimator \eqref{eq:estimator} displays a null systematic error compared to the mathematical probabilistic representation \eqref{eq:Esp}-\eqref{eq:BSP} and the physical model \eqref{eq:Poisson-Vlasov}, and comes with confidence intervals.
Then, this approach is meshless since there is a complete orthogonality between the calculus and the description of geometric space,velocity space and time.
This last remark allow us to affirm that solving problems involving complex geometries yields no conceptual difference nor technical bottleneck, as shown in \cite{Ibarrart_2025, Villefranque_2022,Nyffenegger_2024,Bati_2023}.
Thirdly, this approach allows one to calculate sensitivities from within the main simulation, and parallelization is straightforward.
Finally, this method is a phase-space pointwise method avoiding us from computing the whole distribution function field or having to follow numerous particles interacting which each other.\\

\paragraph{Distribution function path sampling.}

Sampling the random variable $\cc{F}$ implies the ability to construct the path described by the embedded phase-space stochastic process $\{(\boldsymbol{\widetilde{\cc{R}}}_s,\boldsymbol{\widetilde{\cc{C}}}_s)\}_s$. This process is the solution to the system of stochastic differential equations \eqref{eq:BSP} with the initial condition $(\boldsymbol{\cc{R}}_\text{o},\boldsymbol{\cc{C}}_\text{o})=(\mathbf{r},\mathbf{c})$.
As $\boldsymbol{\cc{G}}$ appears itself on the definition of the branching path, it is algorithmically translated by a recursive structure between Alg. \ref{alg:function} and Alg. \ref{alg:force}.
Branching paths are sampled using an Euler discretization scheme corresponding.
Defining $n$ such that $t-t_\text{o}=n\delta s$ and providing us with a regular
subdivision $\{i\delta s|i\in\llbracket0;n-1\rrbracket\}$ of
$[\text{o};t-t_\text{o}]$, this partitioned symplectic (position-first) Euler scheme writes
\begin{equation}
\left\{
\begin{array}{ll}
\delta\boldsymbol{\widehat{\widetilde{\cc{R}}}}_{i\delta s}&=-\boldsymbol{\widehat{\widetilde{\cc{C}}}}_{i\delta s}\delta s\\
\delta\boldsymbol{\widehat{\widetilde{\cc{C}}}}_{i\delta s}&=\Big(\boldsymbol{\cc{G}}\Big|\boldsymbol{\widehat{\widetilde{\cc{R}}}}_{(i+1)\delta s},t-(i+1)\delta s\Big)\delta s/m
\end{array}
\right.\label{eq:BSP_discrete}
\end{equation}
given stochastic increments $\delta\boldsymbol{\widehat{\widetilde{\cc{R}}}}_{i\delta s}=\boldsymbol{\widehat{\widetilde{\cc{R}}}}_{(i+1)\delta s}-\boldsymbol{\widehat{\widetilde{\cc{R}}}}_{i\delta s}$ and $\delta\boldsymbol{\widehat{\widetilde{\cc{C}}}}_{i\delta s}=\boldsymbol{\widehat{\widetilde{\cc{C}}}}_{(i+1)\delta s}-\boldsymbol{\widehat{\widetilde{\cc{C}}}}_{i\delta s}$.
The continuous limit is obtained when $n\to\infty$, that is $\delta s\to0$, and
has to be understood as convergence in probability in the sense of Ito.
In this limit $\{\boldsymbol{\widehat{\widetilde{\cc{R}}}}_s,\boldsymbol{\widehat{\widetilde{\cc{C}}}}_s\}_s$ tend to $\{\boldsymbol{\widetilde{\cc{R}}}_s,\boldsymbol{\widetilde{\cc{C}}}_s\}_s$. 
Alg. \ref{alg:function} presents the sampling method for these random-force branching paths.

\begin{algorithm}[H]
  \caption{$(\cc{F}|\mathbf{r},\mathbf{c},t)$ path sampling}
  \label{alg:function}
  $\bullet$ initial path time: $s=0$\;
   $\bullet $ initial phase point: $(\hat{\mathbf{r}}_s,\hat{\mathbf{c}}_s)=(\mathbf{r},\mathbf{c})$\;
   $\bullet$ path discretization time: $\delta s$\;
   \While{$s< t$}{
   $\bullet$ $\hat{\mathbf{r}}_{s+\delta s}\leftarrow\hat{\mathbf{r}}_s-\mathbf{c}\delta s$\;
   $\bullet$ $s\leftarrow s+\delta s$\;
   $\bullet$ $\mathbf{g}_\phi\leftarrow$ sample $(\boldsymbol{\cc{G}}_\phi\big|\hat{\mathbf{r}}_s,t-s))$ according to Alg. \ref{alg:force}\;
   $\bullet$ $\hat{\mathbf{c}}_s\leftarrow\hat{\mathbf{c}}_{s-\delta s}+\mathbf{g}_\phi\delta s/m$}
   $\bullet$ $s_\text{rand}\leftarrow$ sample $\cc{S}$ according to $p_\cc{S}$\;
   \If{$s_\text{rand}\geq t$}{
   $\bullet$ $\mathfrak{f}\leftarrow f_\text{o}(\hat{\mathbf{r}}_s,\hat{\mathbf{c}}_s)$\;
   }
   \If{$s_\text{rand}<t$}{
   $\bullet$ $b\leftarrow$ sample $\cc{B}(\bb{P}_\text{a})$\;
   \If{$b=1$}{
   $\bullet$ $f\leftarrow f^\star(\hat{\mathbf{r}}_{s_\text{rand}},\hat{\mathbf{c}}_{s_\text{rand}},t-s_\text{rand})$\;
   }
   \If{$b=0$}{
   $\bullet$ $\boldsymbol{\omega}\leftarrow$ sample $\boldsymbol{\Omega'}$ according to $\varphi_{\boldsymbol{\Omega'}}$\;
   $\bullet$ $\hat{\mathbf{c}}_{s_\text{rand}}\leftarrow||\hat{\mathbf{c}}_{s_\text{rand}}||\boldsymbol{\omega'}$\;
   $\bullet$ $\mathfrak{f}\leftarrow$ sample $(\cc{F}|\hat{\mathbf{r}}_{s_\text{rand}},\hat{\mathbf{c}}_{s_\text{rand}},t-s_\text{rand})$\;
   }
   }
   \Return $\mathfrak{f}$
\end{algorithm}
\textcolor{white}{.}
\paragraph{Force field path sampling.}

In the view of random-force ballistic paths occuring between two events described by eqn. \eqref{eq:BSP} and allowing to recover the exact probabilistic representation \eqref{eq:Esp} of $f(\mathbf{r},\mathbf{c},t)$, the sampling procedure of $(\boldsymbol{\cc{G}}_\phi\big|\mathbf{r},t)$ can be read from equation \eqref{eq:grad3}.
Alg. \ref{alg:force} details this procedure. 
Such sampling procedure is exact since the underlaying brownian process $\{\boldsymbol{\cc{R}}_s^\phi\}_s$ is sampled without any bias in the free-space $\bb{R}^3$.
\begin{algorithm}[H]
  \caption{$(\boldsymbol{\cc{G}}_\phi\big|\mathbf{r},t)$ path sampling}
  \label{alg:force}
  $\bullet$ probe position $(\mathbf{r},t)$\;
  $\bullet$ $s^\phi\leftarrow$ sample $\cc{S}^\phi$ according to $p_{\cc{S}^\phi}$\;
  $\bullet$ $\mathbf{c}\leftarrow$ sample $\mathbf{C}$ according to $p_\mathbf{C}$\;
  $\bullet$ $\mathbf{r}_{s^\phi}^\phi\leftarrow$ sample $\boldsymbol{\cc{R}}_{s^\phi}^\phi$ according to $\cc{N}(\mathbf{r},2s^\phi)$\;
  $\bullet$ $\mathfrak{f}\leftarrow$ sample $(\cc{F}|\mathbf{r}_s^\phi,\mathbf{c},t)$ according to Alg. \ref{alg:function}\;
   \Return $\mathbf{g}_\phi=\kappa (\rho_\text{ext}(\mathbf{r}_{s^\phi}^\phi,t)-\mathfrak{f}/p_\mathbf{C}(\mathbf{c}))(\mathbf{r}_{s^\phi}^\phi-\mathbf{r}))/2s^\phi p_\cc{S^\phi}(s^\phi)$
\end{algorithm}
\textcolor{white}{.}
\section{Results and discussions}

\paragraph{Unstationary collisional ion-neutral gas.}

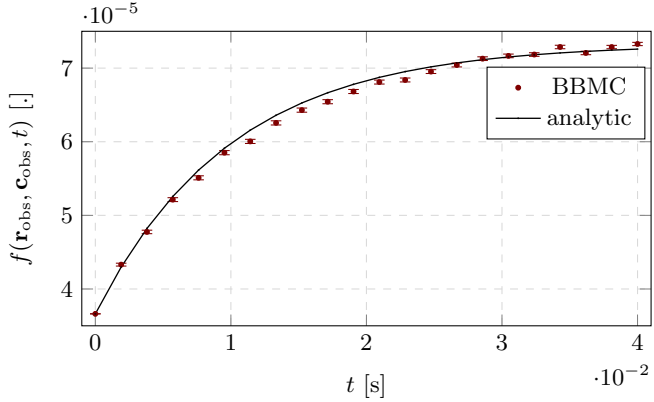
\begin{figure}[H]
\begin{tikzpicture}
		\begin{axis}[width=9.15cm, height=5.5cm, grid=both, grid style={dashed,gray!30}, xlabel={$t$ [s]}, xlabel style={font=\small}, xlabel style={below=-0.0cm}, xtick distance=1e-2, xmin=-1e-3, xmax=4.1e-2, ylabel={$f(\mathbf{r}_\text{obs},\mathbf{c}_\text{obs},t)$ [.]}, ylabel style={font=\small}, ylabel style={above=-0.7cm}, ytick distance=1e-5, ymin=3.5e-5, ymax=7.5e-5, legend style={at={(0.99,0.88)},anchor=north east}]
			
			\addplot[on layer=foreground,
					solid,
					only marks,
					mark=*,
					mark options={scale=0.3},
					line width=1,
					style={black!50!red},
					error bars/.cd,
					y dir = both,
					y explicit] table[x index=0, y index=7, y error index=8] {relaxation_nu.dat};
			\addlegendentry{BBMC};
					
			\addplot[on layer=background,
					solid,
					mark=*,
					mark options={scale=0.04},
					line width=0.5,
					style={black}
					] table[x index=0, y index=9]{relaxation_nu.dat};
  		\addlegendentry{analytic};
  		
		\end{axis}
	\end{tikzpicture}
\caption{Temporal profile of the distribution function at the phase-space probe position $(\mathbf{r}_\text{obs},\mathbf{c}_\text{obs})=(0.01,0.1,0.1,1,10)$. Branching Backward Monte Carlo estimations are computed by use of samples $N=1\times 10^4$ for $\sigma=5\times 10^{-1}$ [m], $\varepsilon_\text{o}=1\times 10^{-3}$ [F.m$^{-1}$], $e=1$ [C], $m=1$ [kg], $\nu_\text{a}=5\times 10^1$ [Hz], $\nu_\text{d}=5\times 10^1$ [Hz], $k_\text{B}=1$, $T=1\times 10^2$ [K], $\delta s=2\times 10^{-3}$ [s] and $\alpha=2$ [m$^{-3}$].}
\label{fig:1}
\end{figure}

\begin{figure}[H]
\begin{tikzpicture}
		\begin{axis}[width=9.15cm, height=6.5cm, grid=both, grid style={dashed,gray!30}, xlabel={$r_{\text{x,obs}}$ [s]}, xlabel style={font=\small}, xlabel style={below=-0.0cm}, xtick distance=1, xmin=-2.6, xmax=2.6, ylabel={$f(\mathbf{r}_\text{obs},\mathbf{c}_\text{obs},t)$ [.]}, ylabel style={font=\small}, ylabel style={above=-0.7cm}, ytick distance=, ymin=-0.1e-5, ymax=6.1e-5, legend style={at={(0.99,0.98)},anchor=north east}]
			
			\addplot[on layer=foreground,
					solid,
					only marks,
					mark=*,
					mark options={scale=0.3},
					line width=1,
					style={black!50!red},
					error bars/.cd,
					y dir = both,
					y explicit] table[x index=0, y index=1, y error index=2] {spatial_1e-3.dat};
			\addlegendentry{BBMC};
					
			\addplot[on layer=background,
					solid,
					mark=*,
					mark options={scale=0.04},
					line width=0.5,
					style={black}
					] table[x index=0, y index=3]{spatial_1e-3.dat};
  		\addlegendentry{analytic};
\addplot[on layer=foreground,
					solid,
					only marks,
					mark=*,
					mark options={scale=0.3},
					line width=1,
					style={black!50!red},
					error bars/.cd,
					y dir = both,
					y explicit] table[x index=0, y index=1, y error index=2] {spatial_5e-3.dat};
					
			\addplot[on layer=background,
					solid,
					mark=*,
					mark options={scale=0.04},
					line width=0.5,
					style={black}
					] table[x index=0, y index=3]{spatial_5e-3.dat};
 \addplot[on layer=foreground,
					solid,
					only marks,
					mark=*,
					mark options={scale=0.3},
					line width=1,
					style={black!50!red},
					error bars/.cd,
					y dir = both,
					y explicit] table[x index=0, y index=1, y error index=2] {spatial_1e-2.dat};
					
			\addplot[on layer=background,
					solid,
					mark=*,
					mark options={scale=0.04},
					line width=0.5,
					style={black}
					] table[x index=0, y index=3]{spatial_1e-2.dat};
  		
		\end{axis}
	\end{tikzpicture}
\caption{Spatial profiles of the distribution function at the phase-space probe position $(\mathbf{r}_\text{obs},\mathbf{c}_\text{obs})=(r_{x,\text{obs}},0.1,0.1,1,10)$. Branching Backward Monte Carlo estimations are computed by use of $N=1\times 10^4$ samples for $\sigma=5\times 10^{-1}$ [m], $\varepsilon_\text{o}=1\times 10^{-3}$ [F.m$^{-1}$], $e=1$ [C], $m=1$ [kg], $\nu_\text{a}=5\times 10^1$ [Hz], $\nu_\text{d}=5\times 10^1$ [Hz], $k_\text{B}=1$, $T=1\times 10^2$, $\alpha=2$ [m$^{-3}$].
Spatial profiles are computed for $t_\text{obs}=1\times 10^{-3}$ [s], $t_\text{obs}=5\times 10^{-3}$ [s] and $t_\text{obs}=1\times 10^{-2}$ [s]}
\label{fig:2}
\end{figure}
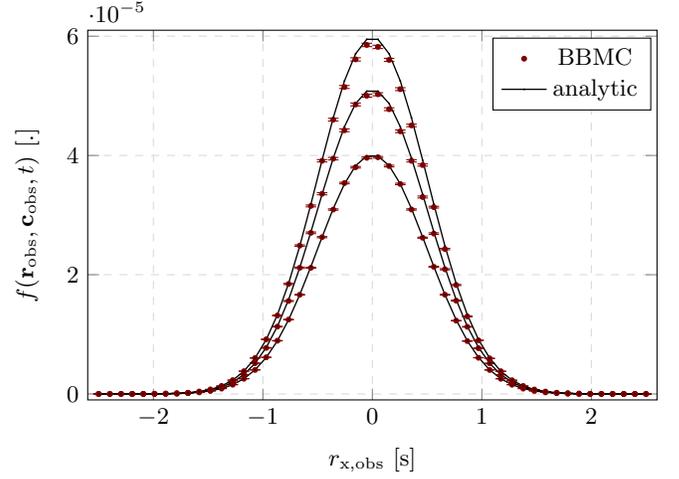

Let us consider a free-space ion gas colliding with a neutral prescribed background.
Ions distribution function $f$ evolve according to \eqref{eq:Poisson-Vlasov} with $\rho_\text{ext}=0$ and $\kappa=-e^2/\varepsilon_\text{o}$.
In the following, an isotropic scatterin phase function $\varphi(\boldsymbol{\omega}|\boldsymbol{\omega'})=1/4\pi$ is choosen.
We fix the initial condition
\begin{equation}
f_{\text{o}}(\mathbf{r},\mathbf{c})=\left(\frac{m}{2\pi k_\text{B}T}\right)^{3/2}\hspace{-0.4cm}\text{exp}\left\{-\frac{m||\mathbf{c}||^2}{2 k_\text{B}T}-\frac{||\mathbf{r}||^2}{2\sigma^2}\right\}(\alpha-1)
\end{equation}
along with the volumic source 
\begin{equation}
\begin{split}
f^\star(\mathbf{r},\mathbf{c},t)=&\left(\frac{m}{2\pi k_\text{B}T}\right)^{3/2}\hspace{-0.4cm}\text{exp}\left\{-\frac{m||\mathbf{c}||^2}{2 k_\text{B}T}-\frac{||\mathbf{r}||^2}{2\sigma^2}\right\}\\
&\hspace{-0.8cm}\times\left(\alpha-\text{exp}\{-\nu_\text{e}t\}\right)\Bigg[1+\frac{\nu_\text{e}}{\nu_\text{a}\left(\alpha\text{e}^{\nu_\text{e}t}-1\right)}-\frac{\mathbf{r}\cdot\mathbf{c}}{\nu_\text{a}\sigma^2}\\
&\hspace{-0.8cm}+\frac{e^2N(t)\mathbf{r}\cdot\mathbf{c}}{4\pi\varepsilon_\text{o}k_\text{B}T||\mathbf{r}||^3}\left(\frac{2\beta}{\sqrt{\pi}}||\mathbf{r}||\text{e}^{-(\beta||\mathbf{r}||)^2}-\text{erf}(\beta||\mathbf{r}||)\right)\Bigg]
\end{split}
\end{equation}
with the number $N(t)=(2\pi \sigma^2)^{3/2}(\alpha-\text{exp}\{-\nu_\text{e}t\})$ of ions and $\beta=(2\sigma^2)^{-1/2}$.
This benchmark allows us to compare our BBMC estimations to the exact analytical distribution function of this initial value problem :
\begin{equation}
\begin{split}
f(\mathbf{r},\mathbf{c},t)=&\left(\frac{m}{2\pi k_\text{B}T}\right)^{3/2}\hspace{-0.4cm}\text{exp}\left\{-\frac{m||\mathbf{c}||^2}{2 k_\text{B}T}-\frac{||\mathbf{r}||^2}{2\sigma^2}\right\}\\
&\times\left(\alpha-\text{exp}\{-\nu_\text{e}t\}\right)
\end{split}
\end{equation}
Fig. \ref{fig:1} and \ref{fig:2} illustrate the comparison between Branching Backward Monte Carlo estimations of the distribution function $f$ at a given probe position as a function of the observation time $t$ and the abscissa.
The full system \eqref{eq:Poisson-Vlasov} consisting in the nonlinear coupling between Boltzmann transport and Poisson's equation is solved by statistical sampling over branching paths within a unique and well-defined path-space.

\paragraph{Plasma relaxation.}

Let us consider now the free-space electron-ion plasma relaxation occuring under charges-neutrals collisions and described by the nonlinearly coupled two-species Boltzmann-Poisson system \eqref{eq:Poisson-Vlasov}, noting $\phi=q\Phi_e$, $q=z_se$ and $z_s=\mathds{1}_{\{s=i\}}-\mathds{1}_{\{s=e\}}$.
\begin{widetext}
\begin{equation}
\left\{
\begin{array}{ll}
\p_tf_s(\mathbf{r},\mathbf{c},t)&\hspace{-0.15cm}+\mathbf{c}\cdot\boldsymbol{\nabla}_\mathbf{r}f_s(\mathbf{r},\mathbf{c},t)-(z_se/m_s)\boldsymbol{\nabla}_\mathbf{r}\Phi(\mathbf{r},t)\cdot\boldsymbol{\nabla}_\mathbf{c}f_s(\mathbf{r},\mathbf{c},t)=-\nu_\text{e}f_s(\mathbf{r},\mathbf{c},t)+\nu_\text{a}f^\star_s(\mathbf{r},\mathbf{c},t)+\nu_\text{d}\int_{\bb{S}^2}\frac{\ds\boldsymbol{\omega}}{4\pi}f_s(\mathbf{r},||\mathbf{c}||\boldsymbol{\omega},t)\\
\nabla^2\Phi(\mathbf{r},t)&=-(e/\varepsilon_\text{o})\int_{\bb{R}^3}\ds\mathbf{c}~\sum_sz_sf_s(\mathbf{r},\mathbf{c},t)\\
f_s(\mathbf{r},\mathbf{c},t_\text{o})&=f_{s,\text{o}}(\mathbf{r},\mathbf{c})
\end{array}
\right.
\end{equation}
\end{widetext}
The volumic source $f^\star$ and the initial condition $f_\text{o}$ are expressed in App. \ref{app:1}.
Given these conditions, the solution $f_s$ of this coupled systems writes as follows: 
\begin{equation}
f_s(\mathbf{r},\mathbf{c},t)=\rho_s(\mathbf{r},t)\left(\frac{m_s}{2\pi k_\text{B}T_s}\right)^{3/2}\hspace{-0.4cm}\text{exp}\left\{-\frac{m_s||\mathbf{c}-\mathbf{v}_s(\mathbf{r},t)||^2}{2k_\text{B}T_s}\right\}
\end{equation}
given
\begin{equation}
\begin{split}
\rho_s(\mathbf{r},t)&=\int_{\bb{R}^3}\ds\mathbf{c}~f_s(\mathbf{r},\mathbf{c},t)\\
&=\rho_\infty+\rho_\text{o}\text{exp}\left\{-\frac{||\mathbf{r}||^2}{2\sigma_s^2}\right\}\text{exp}\{-\nu_\text{e}t\}
\end{split}
\end{equation}
and
\begin{equation}
\begin{split}
\mathbf{v}_s(\mathbf{r},t)&=\frac{1}{\rho_s(\mathbf{r},t)}\int_{\bb{R}^3}\ds\mathbf{c}~\mathbf{c}f_s(\mathbf{r},\mathbf{c},t)\\
&=\frac{-z_s\nu_\text{e}\mathbf{r}}{4\pi e||\mathbf{r}||^3\sum_s\rho_s(\mathbf{r},t)}\\
&\times\sum_sz_sQ_s(t)\left(\frac{2\beta_s}{\sqrt{\pi}}\text{e}^{-(\beta_s||\mathbf{r}||)^2}-\text{erf}\left(\beta_s||\mathbf{r}||\right)\right)
\end{split}
\end{equation}
in which the total number of particles $N_s(t)=\rho_\text{o}(2\pi\sigma_s^2)^{3/2}\text{exp}\{-\nu_\text{e}t\}$, the total charge $Q_s(t)=eN_s(t)$ and $\beta_s=(2\sigma_s^2)^{-1}$.
It can be shown (see App. \ref{app:1}) that such solution satisfies the ionization detailed balance constraint
\begin{equation}
\int_{\bb{R}^3}\ds\mathbf{c}~\sum_sz_sef_s(\mathbf{r},\mathbf{c},t)=\int_{\bb{R}^3}\ds\mathbf{c}~\sum_sz_sef^\star_s(\mathbf{r},\mathbf{c},t)
\end{equation}
being in total accordance with charge conservation:
\begin{equation}
\p_t\left(\sum_sz_se\rho_s(\mathbf{r},t)\right)=-\boldsymbol{\nabla}\cdot\left(\sum_sz_se\rho_s(\mathbf{r},t)\mathbf{v}_s(\mathbf{r},t)\right)
\end{equation}

\begin{figure}[H]
\begin{tikzpicture}
		\begin{axis}[width=9.15cm, height=5.5cm, grid=both, grid style={dashed,gray!30}, xlabel={$t$ [s]}, xlabel style={font=\small}, xlabel style={below=-0.0cm}, xtick distance=1e-6, xmin=-0.3e-6, xmax=6.1e-6, ylabel={$f(\mathbf{r}_\text{obs},\mathbf{c}_\text{obs},t)$ [.]}, ylabel style={font=\small}, ylabel style={above=-0.7cm}, ytick distance=, ymin=5e-10, ymax=4.3e-9, legend style={at={(0.99,0.92)},anchor=north east}]
			
			\addplot[on layer=foreground,
					solid,
					only marks,
					mark=*,
					mark options={scale=0.3},
					line width=1,
					style={black!50!red},
					error bars/.cd,
					y dir = both,
					y explicit] table[x index=0, y index=1, y error index=2] {relaxation.dat};
			\addlegendentry{BBMC};
					
			\addplot[on layer=background,
					solid,
					mark=*,
					mark options={scale=0.04},
					line width=0.5,
					style={black}
					] table[x index=0, y index=3]{relaxation.dat};
  		\addlegendentry{analytic};
  		
		\end{axis}
	\end{tikzpicture}
\caption{Temporal profile of the electron distribution function $f_e$ at the phase-space probe position $(\mathbf{r}_\text{obs},\mathbf{c}_\text{obs})=(1\times 10^{-2},1\times 10^{-2},1\times 10^{-1},1\times 10^{2},1\times 10^{2},1)$. Branching Backward Monte Carlo estimations are computed by use of $N=1\times 10^4$ samples for $\sigma_i=1\times 10^{-4}$ [m], $\rho_\text{o}=1\times 10^{11}$ [m$^{-3}$], $\sigma_e=1\times 10^{-1}$ [m], $T_i=0.05\times1,160\times 10^4$ [K], $T_e=6\times 1,160\times 10^4$ [K], $k_\text{B}=1.380\times 10^{-23}$, $\varepsilon_\text{o}=8,854\times 10^{-12}$ [F.m$^{-1}$], $e=1,602\times 10^{-19}$ [C], $m_i=6,6\times 10^{-26}$ [kg], $m_e=9,109\times 10^{-31}$ [kg], $\delta s=7\times 10^{-7}$ [s], $\nu_\text{d}=1\times 10^5$ [Hz], $\nu_\text{a}=4\times 10^5$ [Hz], $\rho_\infty=1\times 10^{10}$.}
\label{fig:3}
\end{figure}
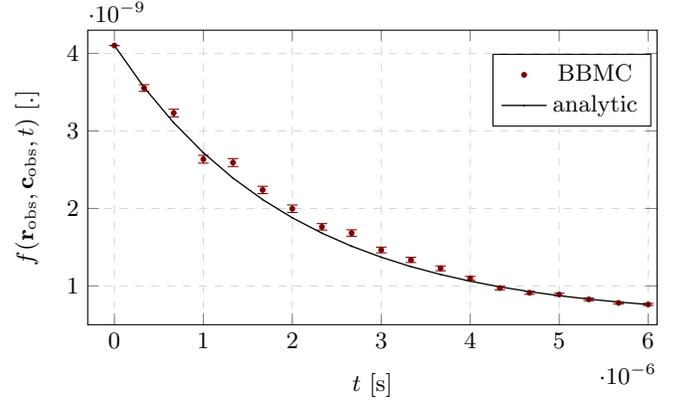

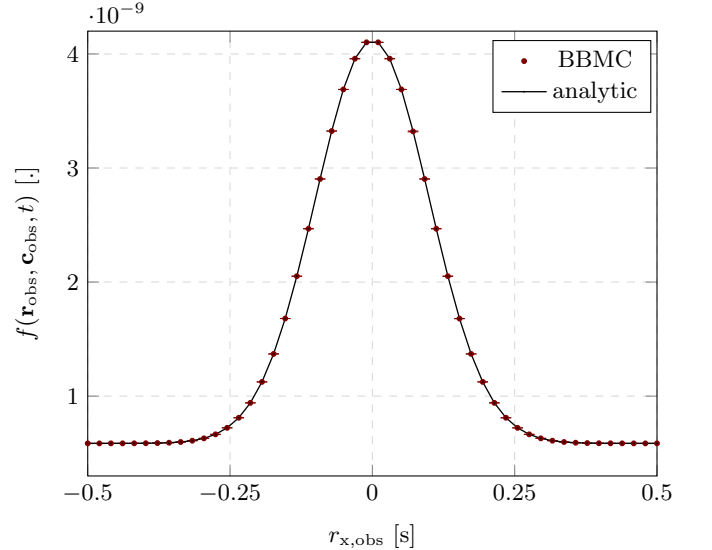
\begin{figure}[H]
\begin{tikzpicture}
		\begin{axis}[width=9.15cm, height=7.5cm, grid=both, grid style={dashed,gray!30}, xlabel={$r_{\text{x,obs}}$ [s]}, xlabel style={font=\small}, xlabel style={below=-0.0cm}, xtick distance=0.25, xmin=-0.5, xmax=0.5, ylabel={$f(\mathbf{r}_\text{obs},\mathbf{c}_\text{obs},t)$ [.]}, ylabel style={font=\small}, ylabel style={above=-0.7cm}, ytick distance=, ymin=3e-10, ymax=4.2e-9, legend style={at={(0.99,0.98)},anchor=north east}]
			
			\addplot[on layer=foreground,
					solid,
					only marks,
					mark=*,
					mark options={scale=0.3},
					line width=1,
					style={black!50!red},
					error bars/.cd,
					y dir = both,
					y explicit] table[x index=0, y index=1, y error index=2] {spatial_8e-8.dat};
			\addlegendentry{BBMC};
					
			\addplot[on layer=background,
					solid,
					mark=*,
					mark options={scale=0.04},
					line width=0.5,
					style={black}
					] table[x index=0, y index=3]{spatial_8e-8.dat};
  		\addlegendentry{analytic};
  		
		\end{axis}
	\end{tikzpicture}
\caption{Spatial profile of the electron distribution function $f_e$ at the phase-space probe position $(\mathbf{r}_\text{obs},\mathbf{c}_\text{obs})=(r_{x,\text{obs}},1\times 10^{-2},1\times 10^{-1},1\times 10^{2},1\times 10^{2},1)$. Branching Backward Monte Carlo estimations are computed by use of $N=1\times 10^4$ samples for $\sigma_i=1\times 10^{-4}$ [m], $\rho_\text{o}=1\times 10^{11}$ [m$^{-3}$], $\sigma_e=1\times 10^{-1}$ [m], $T_i=0.05\times1,160\times 10^4$ [K], $T_e=6\times 1,160\times 10^4$ [K], $k_\text{B}=1.380\times 10^{-23}$, $\varepsilon_\text{o}=8,854\times 10^{-12}$ [F.m$^{-1}$], $e=1,602\times 10^{-19}$ [C], $m_i=6,6\times 10^{-26}$ [kg], $m_e=9,109\times 10^{-31}$ [kg], $\delta s=3\times 10^{-8}$ [s], $\nu_\text{d}=1\times 10^5$ [Hz], $\nu_\text{a}=4\times 10^5$ [Hz], $\rho_\infty=1\times 10^{10}$, and $t_\text{obs}=3\times 10^{-7}$ [s].}
\label{fig:4}
\end{figure}

Results illustrated in Fig. \ref{fig:3} and \ref{fig:4} provide us with robust comparison between analytic solution of the nonlinearly coupled Poisson-Boltzmann system and Branching Backward Monte Carlo estimations sampling a unique branching path-space.


\section{conclusion}

In the present work, we have advanced recent probabilistic approaches of nonlinear advecto-reacto-diffusive transport to the class of mesoscopic Boltzmann transport models nonlinearly coupled to Poisson's submodels of the self-consistent force-field, while enabling explicit propagator representations.
Our formulation shows how expectations over a single, well-defined branching path-space recover analytical solutions, as schown in plasma physics.
Taken together, these results bridge physical interpretation and computational feasibility across scientific communities concened with nonlinear mesoscopic transport phenomena and offer a new descriptive framework, one capable of honoring the full complexity of the underlying physics while delivering a tractable, insightful representation.


\begin{acknowledgments}
This work was supported by 
the MCMET project (ANR-23-CE46-0002) 
of the French National Research Agency
(ANR).
\end{acknowledgments}

\bibliography{biblio_merge}
\clearpage
\appendix
\section{Analytical solutions for the relaxing plasma configuration.}\label{app:1}

In this appendix, we detail the analytical sketch aiming at founding an exact solution $f_s$, $s\in\{e,i\}$, to the free-space coupled Poisson-linear Boltzmann equation
\begin{widetext}
\begin{equation}
\left\{
\begin{array}{ll}
\p_tf_s(\mathbf{r},\mathbf{c},t)&\hspace{-0.15cm}+\mathbf{c}\cdot\boldsymbol{\nabla}_\mathbf{r}f_s(\mathbf{r},\mathbf{c},t)-(z_se/m_s)\boldsymbol{\nabla}_\mathbf{r}\Phi(\mathbf{r},t)\cdot\boldsymbol{\nabla}_\mathbf{c}f_s(\mathbf{r},\mathbf{c},t)=-\nu_\text{e}f_s(\mathbf{r},\mathbf{c},t)+\nu_\text{a}f^\star_s(\mathbf{r},\mathbf{c},t)+\nu_\text{d}\int_{\bb{S}^2}\frac{\ds\boldsymbol{\omega}}{4\pi}f_s(\mathbf{r},||\mathbf{c}||\boldsymbol{\omega},t)\\
\nabla^2\Phi(\mathbf{r},t)&=-(e/\varepsilon_\text{o})\int_{\bb{R}^3}\ds\mathbf{c}~\sum_sz_sf_s(\mathbf{r},\mathbf{c},t)\\
f_s(\mathbf{r},\mathbf{c},t_\text{o})&=f_{s,\text{o}}(\mathbf{r},\mathbf{c})
\end{array}
\right.\label{eq:app:0}
\end{equation}
\end{widetext}
fulfilling the detailed balance constraint
\begin{equation}
\int_{\bb{R}^3}\ds\mathbf{c}~\sum_sz_sef_s(\mathbf{r},\mathbf{c},t)=\int_{\bb{R}^3}\ds\mathbf{c}~\sum_sz_sef^\star_s(\mathbf{r},\mathbf{c},t)\label{eq:app:1}
\end{equation}
The idea is to backwardly design $f_s^\star$ for \eqref{eq:app:1} to be verified.\\

\paragraph{Sketch of the proof.}

The very first idea is to fix the functional expression of ions and electrons distribution function. The choice that have been made is a Maxwellian ditribution shifted by the field $\mathbf{v}_s$ which will be determined afterall by the ionization balance constraint. 
\begin{equation}
f_s(\mathbf{r},\mathbf{c},t)=\rho_s(\mathbf{r},t)\left(\frac{m_s}{2\pi k_\text{B}T_s}\right)^{3/2}\hspace{-0.45cm}\text{exp}\left\{-\frac{m_s||\mathbf{c}-\mathbf{v}_s(\mathbf{r},t)||^2}{2k_\text{B}T_s}\right\}
\end{equation}
The density field is choosen as the sum of a uniform field $\rho_\infty$ and a Gaussian spatial density exponentially attenuated in time : 
\begin{equation}
\rho_s(\mathbf{r},t)=\rho_\infty+\rho_\text{o}\text{exp}\left\{-\frac{||\mathbf{r}||^2}{2\sigma_s^2}\right\}\text{exp}\{-\nu_\text{e}t\}
\end{equation}
Then, as the electric potential $\Phi$ is solution of Poisson's equation $\nabla^2\Phi=-(e/\varepsilon_\text{o})\sum_sz_s\rho_s$, one can deduce its expression from the convolution between the Green kernel of Poisson's equation and its sources :  
\begin{equation}
\begin{split}
\Phi(\mathbf{r},t)&=\frac{e}{4\pi\varepsilon_\text{o}}\int_{\bb{R}^3}\ds\mathbf{r'}~\frac{1}{||\mathbf{r}-\mathbf{r'}||}\sum_sz_s\rho_s(\mathbf{r'},t)\\
&=\frac{1}{4\pi\varepsilon_\text{o}||\mathbf{r}||}\sum_sz_sQ_s(t)\text{erf}\left(\beta_s||\mathbf{r}||\right)
\end{split}
\end{equation}
given the number of particles $s$, $N_s(t)=\rho_\text{o}(2\pi\sigma_s^2)^{3/2}\text{exp}\{-\nu_\text{e}t\}$, the total charge of particles $s$, $Q_s(t)=eN_s(t)$ and $\beta_s=(2\sigma_s^2)^{-1/2}$.
The integration is straightforward since $\rho_s$ have been choosen, in this view, as a Gaussian spatial density.
Therefore, $f_s^\star$ have to be defined as 
\begin{widetext}
\begin{equation}
f_s^\star(\mathbf{r},\mathbf{c},t)=\frac{1}{\nu_\text{a}}\left(\p_tf_s(\mathbf{r},\mathbf{c},t)+\mathbf{c}\cdot\boldsymbol{\nabla}_\mathbf{r}f_s(\mathbf{r},\mathbf{c},t)\right)-\frac{z_se}{\nu_\text{a}m_s}\boldsymbol{\nabla}_\mathbf{r}\Phi(\mathbf{r},t)\cdot\boldsymbol{\nabla}_\mathbf{c}f_s(\mathbf{r},\mathbf{c},t)+\frac{\nu_\text{e}}{\nu_\text{a}}f_s(\mathbf{r},\mathbf{c},t)-\frac{\nu_\text{d}}{\nu_\text{a}}\int_{\bb{S}^2}\frac{\ds\boldsymbol{\omega}}{4\pi}f_s(\mathbf{r},||\mathbf{c}||\boldsymbol{\omega},t)\label{ici}
\end{equation}
\end{widetext}
in accordance with \eqref{eq:app:0}.
The first transport term admits an exact analytical expression involving spatial and temporal derivatives of $\mathbf{v}_s$ which are expressed in the last paragraph of this appendix :
\begin{widetext}
\begin{equation}
\begin{split}
\p_tf_s(\mathbf{r},\mathbf{c},t)+\mathbf{c}\cdot\boldsymbol{\nabla}_\mathbf{r}f_s(\mathbf{r},\mathbf{c},t)=&-\frac{f_s(\mathbf{r},\mathbf{c},t)}{\sum_s\rho_s(\mathbf{r},t)}(\rho_s(\mathbf{r},t)-\rho_\infty)\left(\nu_\text{e}+\frac{\mathbf{c}\cdot\mathbf{r}}{\sigma_s^2}\right)\\
&\hspace{-3.5cm}+\frac{m_s}{k_\text{B}T_s}f_s(\mathbf{r},\mathbf{c},t)\left[\left(\frac{\mathbf{c}\cdot\mathbf{r}}{||\mathbf{r}||}-||\mathbf{v}_s(\mathbf{r},t)||\right)\left(\p_t||\mathbf{v}_s(\mathbf{r},t)||+\frac{\mathbf{c}\cdot\mathbf{r}}{||\mathbf{r}||}\p_{||\mathbf{r}||}||\mathbf{v}_s(\mathbf{r},t)||\right)+\frac{||\mathbf{v}_s(\mathbf{r},t)||}{||\mathbf{r}||}\left(||\mathbf{c}||^2-\left(\frac{\mathbf{c}\cdot\mathbf{r}}{||\mathbf{r}||}\right)^2\right)\right]
\end{split}
\end{equation}
\end{widetext}
Then, the second transport term appearing in \eqref{ici} involve on one hand the velocity gradient of the distribution function
\begin{equation}
\boldsymbol{\nabla}_\mathbf{c}f_s(\mathbf{r},\mathbf{c},t)=-\frac{m_s}{k_\text{B}T_s}(\mathbf{c}-\mathbf{v}_s(\mathbf{r},t))f_s(\mathbf{r},\mathbf{c},t)
\end{equation}
and on the other hand, the spatial gradient of the electric potential
\begin{equation}
\boldsymbol{\nabla}\Phi(\mathbf{r},t)=\frac{\mathbf{r}}{4\pi\varepsilon_\text{o}||\mathbf{r}||^3}\sum_sz_sQ_s(t)\Gamma_s(\mathbf{r})
\end{equation}
given
\begin{equation}
\Gamma_s(\mathbf{r})=\frac{2\beta_s}{\sqrt{\pi}}||\mathbf{r}||\text{e}^{-(\beta_s||\mathbf{r}||)^2}-\text{erf}\left(\beta_s||\mathbf{r}||\right)
\end{equation}
Finally, it can be shown that the scattering integral also admits an exact analytical expression : 
\begin{widetext}
\begin{equation}
\int_{\bb{S}^2}\frac{\ds\boldsymbol{\omega}}{4\pi}f_s(\mathbf{r},||\mathbf{c}||\boldsymbol{\omega},t)=\rho_s(\mathbf{r},t)\left(\frac{m_s}{2\pi k_\text{B}T_s}\right)^{3/2}\text{exp}\left\{-\frac{m_s(||\mathbf{c}||^2+||\mathbf{v}_s(\mathbf{r},t)||^2)}{2k_\text{B}T_s}\right\}\text{sinhc}\left(\frac{m_s||\mathbf{c}||||\mathbf{v}_s(\mathbf{r},t)||}{k_\text{B}T_s}\right)
\end{equation}
\end{widetext}
noting sinhc$(x)$=sinh$(x)/x$ the hyperbolic cardinal sinus function.\\

\paragraph{Ionization balance and charge conservation.}

The Boltzmann transport equation \eqref{eq:app:0} can be multiplied by $z_se$ and summed over $s$ before being integrated over the velocity space. In doing so, one gets 
\begin{equation}
\begin{split}
\p_t\left(\sum_sz_se\rho_s(\mathbf{r},t)\right)=&-\boldsymbol{\nabla}\cdot\left(\sum_sz_se\rho_s(\mathbf{r},t)\mathbf{v}_s(\mathbf{r},t)\right)\\
&\hspace{-2cm}+\int_{\bb{R}^3}\ds\mathbf{c}\sum_sz_se(-\nu_\text{e}f_s(\mathbf{r},\mathbf{c},t)+\nu_\text{a}f_s^\star(\mathbf{r},\mathbf{c},t))\\
&\hspace{-2cm}+\int_{\bb{R}^3}\ds\mathbf{c}~\sum_sz_se\nu_\text{d}\int_{\bb{S}^2}\frac{\ds\boldsymbol{\omega}}{4\pi}f_s(\mathbf{r},||\mathbf{c}||\boldsymbol{\omega},t)
\end{split}
\end{equation}
However, one can show that
\begin{equation}
\int_{\bb{R}^3}\ds\mathbf{c}~\int_{\bb{S}^2}\frac{\ds\boldsymbol{\omega}}{4\pi}f_s(\mathbf{r},||\mathbf{c}||\boldsymbol{\omega},t)=\int_{\bb{R}^3}\ds\mathbf{c}~f_s(\mathbf{r},\mathbf{c},t)
\end{equation}
Hence, charge conservation
\begin{equation}
\p_t\left(\sum_sz_se\rho_s(\mathbf{r},t)\right)=-\boldsymbol{\nabla}\cdot\left(\sum_sz_se\rho_s(\mathbf{r},t)\mathbf{v}_s(\mathbf{r},t)\right)
\end{equation}
holds if the detailed balance condition
\begin{equation}
\int_{\bb{R}^3}\ds\mathbf{c}~\sum_sz_sef_s(\mathbf{r},\mathbf{c},t)=\int_{\bb{R}^3}\ds\mathbf{c}~\sum_sz_sef^\star_s(\mathbf{r},\mathbf{c},t)
\end{equation}
is satisfied, since $\nu_\text{e}=\nu_\text{a}+\nu_\text{d}$.\\

\paragraph{Functional form of $\mathbf{v}_s$.}

One can show that $\mathbf{v}_s$ has to be written
\begin{equation}
\mathbf{v}_s(\mathbf{r},t)=-\frac{z_s\nu_\text{e}\varepsilon_\text{o}}{e\sum_s\rho_s(\mathbf{r},t)}\boldsymbol{\nabla}\Phi(\mathbf{r},t)
\end{equation}
Indeed, on one hand 
\begin{equation}
\p_t\left(\sum_sz_se\rho_s(\mathbf{r},t)\right)=-\nu_\text{e}\sum_sz_se\rho_s(\mathbf{r},t)
\end{equation}
and on the other hand,
\begin{equation}
\begin{split}
\boldsymbol{\nabla}\cdot\left(\sum_sz_se\rho_s(\mathbf{r},t)\mathbf{v}_s(\mathbf{r},t)\right)&=\boldsymbol{\nabla}\cdot\left(-\nu_\text{e}\varepsilon_\text{o}\boldsymbol{\nabla}\Phi(\mathbf{r},t)\right)\\
&=-\nu_\text{e}\varepsilon_\text{o}\nabla^2\Phi(\mathbf{r},t)\\
&=\nu_\text{e}\sum_sz_se\rho_s(\mathbf{r},t)
\end{split}
\end{equation}
since Poisson's equation is satisfied by $\Phi$.
Hence
\begin{equation}
\mathbf{v}_s(\mathbf{r},t)=-\frac{z_s\nu_\text{e}}{4\pi e\sum_s\rho_s(\mathbf{r},t)}\frac{\mathbf{r}}{||\mathbf{r}||^3}\sum_sz_sQ_s(t)\Gamma_s(\mathbf{r})
\end{equation}
\textcolor{white}{.}

\paragraph{Spatial and temporal derivatives of $||\mathbf{v}_s||$.}

Since the previous detailed balance constraint imposes an expression of $\mathbf{v}_s$, one can now express the exact spatial and temporal derivatives of $||\mathbf{v}_s||$, as they are involved in the formal expression of $f^\star_s$.
One can show that the temporal derivative is given by
\begin{equation}
\p_t||\mathbf{v}_s(\mathbf{r},t)||=\frac{z_s\nu_\text{e}^2\rho_\infty}{2\pi e\sum_s\rho_s(\mathbf{r},t)}\sum_sz_sQ_s(t)\frac{\Gamma_s(\mathbf{r})}{||\mathbf{r}||^2}
\end{equation}
Concerning, the spatial derivative, its expression follows
\begin{equation}
\begin{split}
\p_{||\mathbf{r}||}||\mathbf{v}_s(\mathbf{r},t)||=&\frac{-\nu_\text{e}z_s}{4\pi e\sum_s\rho_s(\mathbf{r},t)}\sum_sz_sQ_s(t)\left(\frac{\text{e}^{-(\beta_s||\mathbf{r}||)^2}}{\sigma_s^3\sqrt{2\pi}}-2\frac{\Gamma_s(\mathbf{r})}{||\mathbf{r}||^3}\right)\\
&+\frac{\nu_\text{e}z_s}{4\pi e\left(\sum_s\rho_s(\mathbf{r},t)\right)^2}\xi(\mathbf{r},t)\sum_sz_sQ_s(t)\frac{\Gamma_s(\mathbf{r})}{||\mathbf{r}||^2}
\end{split}
\end{equation}
with
\begin{equation}
\xi(\mathbf{r},t)=-\rho_\text{o}||\mathbf{r}||\text{e}^{-\nu_\text{e}t}\sum_s\frac{\text{e}^{-(\beta_s||\mathbf{r}||)^2}}{\sigma_s^2}
\end{equation}

\end{document}